# Comprehensiveness of Archives:
## A Modern AI-enabled Approach to Build Comprehensive Shared Cultural Heritage

**Abhishek Gupta[1,2] and Nikitasha Kapoor[3]**

[1]Founder, Montreal AI Ethics Institute
[2]Machine Learning Engineer, Microsoft
[3]Principal Researcher, Pure & Applied Group

**Correspondence**
Email: abhishek@montrealethics.ai; nikitasha@pureandapplied.group.

10 April 2020, 23:50 EST.

## Abstract

Archives play a crucial role in the construction and advancement of society. Humans place a great deal of trust in archives and depend on them to craft public policies and to preserve languages, cultures, self-identity, views and values. Yet, there are certain voices and viewpoints that remain elusive in the current processes deployed in the classification and discoverability of records and archives.

In this paper, we explore the ramifications and effects of centralized, due process archival systems on marginalized communities. There is strong evidence to prove the need for progressive design and technological innovation while in the pursuit of comprehensiveness, equity and justice.

Intentionality and comprehensiveness is our greatest opportunity when it comes to improving archival practices and for the advancement and thrive-ability of societies at large today. Intentionality and comprehensiveness is achievable with the support of technology and the Information Age we live in today. Reopening, questioning and/or purposefully including others voices in archival processes is the intention we present in our paper.

We provide examples of marginalized communities who continue to lead "community archive" movements in efforts to reclaim and protect their cultural identity, knowledge, views and futures. In conclusion, we offer design and AI-dominant technological considerations worth further investigation in efforts to bridge systemic gaps and build robust archival processes.

---

## 1. What is the significance we, as a society, give to archives today?

There is a strong presence of trust and natural reliance placed in archives today (Brown, 2019). While the neuroscience research around "trust" is under-developed, it is completely human for us to trust in systems like archives – a centralized and accessible directory of records. Humans have an innate capacity to trust and cooperate with natural or man-made systems when believed to be vital to the advancement and thriveability of communities and nations (Faulkner and Simpson, 2017).

Archives are an accumulation of recorded information of an organization, community, human and/or nation. Archives have a crucial role in the construction of our social memory (Blouin, 1999). Archives adapt and withstand time with the leadership of archivists and archive organizations. Today, archivists and archive organizations play a custodian decision-making and advisory role in how records are classified and stored; appraisal and disposition are common processes in order to research and define archival value (International Council on Archives, 2016). It thus becomes vital to ensure archivists have the necessary knowledge, skills, tools and comprehension to classify and weigh the significance of a recorded artifact for one social group versus another.

## 2. How can society benefit from recent movements in archival practices?

Throughout history, the users of records are not those who created them. Records and archives can be used hundreds of years after the records were created and that is because the contextual value and perceived weight of value grows overtime. Similar to how businesses rely on historics to base future decisions, societies have time over time relied on preceding choices, biases, stereotypes, values and wisdoms from the past (Braun et al., 2018).

We are interested in exploring the ramifications of centralized, due process archival systems on marginalized communities. There is strong evidence to prove the need for progressive design and technological changes for the pursuit of comprehensiveness, equity and justice.

With admirable eloquence, Flinn highlights the powerful role independent "community archive" movements play in adding the needed marginalized voices and views to traditionally dominant records and archives. According to Flinn, the term "community archives" can have different meanings and interpretations. The broadly accepted definitions tend to be "focused on the activity rather than the form [...] community archives and heritage initiatives come in many different forms and seek to document the history of all manner of local, occupational, ethnic, faith and other diverse communities. They do so by collecting, preserving and making accessible documents, photographs, oral histories and many other materials which document the histories of particular groups and localities" (2010, p 146).

The collective originality, creativity and stewardship of individuals and communities has empowered our individuals to question and improve the comprehensiveness of existing archives. The grassroot movement of community archives has created a profound space for individuals and marginalized communities to archive for the sake of self-identification, community cohesion, community health, and holistic policy development (2010, p. 145). We, as authors of this position paper, are encouraged to further explore how design interventions and technology can support in lessening the barriers and amplify the usefulness of preserving archives and community archives.

## 2.1 Leading with Intentionality

Female archeologists, anthropologists, social scientists, authors and many other leaders firmly identify the incomplete and harmful effects of male-dominant archives of women in society. The role of women has been primarily written by men, and often written in "the realms of sex, religion, custom, culture, politics and economics" (Armstrong, 2019). This very reliance on male-dominant archival women in society is what leads to sexist, incomplete, harmful decisions in how we advance societies today.

Indigenous communities continue to fight for their rights, identity and worldviews. The colonial oppression and the Western-dominate worldviews and archives have endangered and devastated communities of their holistic sense of health, self and happiness (Alexiuk, 2013; Diochon, 2013; MacArthur and Rassmussen, 2017). There are many biases in current archives, often stemming from privilege (Grout, 2019) of who had the ability to create them and thus imbue their lens and perspective that is often seen as the historical truth. These biases are particularly problematic when history can be distorted to portray the viewpoint of one over the other; indigenous peoples often have to document and prove their rights over what has been documented by their oppressors (Association of Canadian Archivists, 2007). The oppression extends to even current practices in archiving when it is done by those who are non-indigenous.

Recent generations and Aboriginal scholars are paving the way to protect their Traditional Knowledge, culture, values and ways of life. Justice can only be achieved with reconciliation and with Indigenous-led rebuilding of our understanding of Indigenous life and culture. The Association of Canadian Archivists state the importance of community archives and community ownership of archives:

> "The new generation of Aboriginal scholars has begun to give more weight to the written word. The fight for the rights of Aboriginal peoples requires that all their records be well preserved, and that the job of Keeper of the Record becomes more important. Recorded words and histories are to defend collective rights in courts and to strengthen and share collective memories in classrooms and history books, in newspapers and photographs, on tape recordings, and on every type of magnetic medium and digital device" (2007).

Intentionality and comprehensiveness is our greatest opportunity when it comes to improving archival practices and for the advancement and thrive-ability of societies at large today. Intentionality and comprehensiveness is achievable with the support of technology and the Information Age we live in today. Reopening, questioning and/or purposefully including others voices in archival processes is the intention we present in our paper.

## 3. The technological challenge and opportunity

While the internet has enabled a larger populace to self-document and own their own narratives, such records continue to appear online in fragmented ways – leading to low discoverability and digital marginalization. The collation efforts by archivists are thus limited by the technical ability to find fragmented context-rich records. Automated methods – web crawlers, discovery algorithms and other online nudge approaches – create real challenges for discoverability of lesser dominant records. Records become especially obsolete by online search engines and tools when they are not in the dominant languages of the internet, for example.

Manual methods of search, discovery and subsequent archiving though quickly go out of sync with the enormous pace of cultural evolution (Perreault, 2012) which outpaces even technological evolution. Archival practices need to be able to capture as much as possible, in as neutral a way as possible. This process is critical to make artifacts available to future knowledge seekers so that they can interpret them in the context and culture of their time.

We have identified several areas where current archival practices fall short in serving the needs of marginalized populations online and offline: Indigenous peoples, women, children, LGBTQIA2+, senior citizens, victims of genocides, racial minorities, cultural minories, military veterans, hearing, visually, and physically challenged persons and more.

The overvalued role of centralized archives and archival organizations is counter-productive to our society's need for a shared culture and equitable governance. Vocal minorities continue to be less discoverable online and in part due to skews in the automated archiving process towards a biased and narrow subset of content creators who know how to gamify online algorithms and increase their content's visibility online. This skew in content discoverability has dramatic implications for what the systems identify high-value archives (Mustafaraj et al., 2011).

Systems that do not have human or community-led supervision pose challenges in how records are interpreted and classified. Artificial Intelligence (AI) systems can systematically improve discoverability and comprehensiveness by systematically maximizing the diversity of viewpoints. Comprehensiveness can be achieved by training AI systems to scour for content that goes beyond what indexes well on the internet. Additionally, AI systems can be leveraged to enhance discoverability of low-traffic, fragmented pockets where some of the smaller communities self-document. While current AI systems have limitations in terms of diversity of languages to be able to parse this kind of content, this is quickly being expanded by efforts of both technical and social science communities from around the world to build better machine language translation systems (Devlin, 2019; ICLR, 2020). Benefits of higher discoverability do not only accrue to marginalized communities; they also create positive knock-on effects for others who gain a better understanding of these cultures and are thus able to truly appreciate our shared cultural heritage in its entirety. On the subject of comprehensiveness, collation of content from automated systems will enhance the available corpus in the archives thus providing a fuller picture to those seeking to build a better understanding of minority cultures.

As individuals increasingly self-identify as digital natives, Archivists, archive organizations and independent archive communities have the opportunity to develop tools and design interventions to more meaningfully connect with contextually-rich records available online. We can improve our engagement with dominant and lesser dominant records online. Hypothetically, an AI-enabled approach to build systems or chatbots to interact with knowledge seekers via commonly used messaging interfaces can lead to increasing their ability to discover a lesser dominant, equally relevant archived artifacts. Such

AI-enabled mechanisms can enhance interactivity and can become a tool for gaining insights from both dominant and lesser dominant sources of content and knowledge; it can also allow individuals to develop digital literacy skills, be exposed to diverse perspectives on historical artifacts and identify misinformation. Such AI-enabled interfaces can also leverage the concept of micro-tasks (Hahn et al., 2019) that helps to break up traditional knowledge into digestible chunks making them more accessible to both serious and casual audiences. Other promising design interventions can include presenting varied perspectives in a web interface side-by-side rather than as ordered list which exhibits cardinality which doesn't make much sense when it comes to understanding, as an example, traditional indigenous artifacts that are subject to interpretations based on the perspective of the observer.

## 4. Final remarks

In this position paper, we have covered some gaps present in archiving processes today that disproportionately impact minorities, not just from misrepresentation, but also from policy and other decisions that are taken based on these incomprehensive, unintentional and non-inclusive archives. We find that modern AI-enabled approaches can bridge some of these gaps creating wider participation in shaping our shared cultural heritage while empowering minority communities to have greater control over knowledge and artifacts that serve to represent their past and shape their present and future identities. We invite other researchers and archivists to build upon the findings here to build archives that are truly inclusive, comprehensive and intentional.

## About the Authors

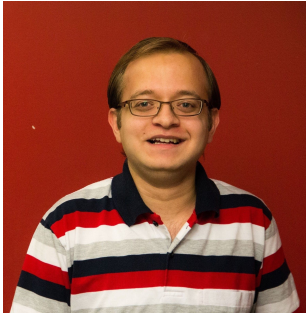

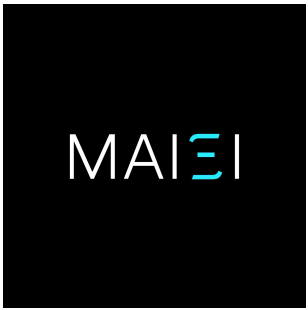

**Abhishek Gupta** is the founder of Montreal AI Ethics Institute (https://montrealethics.ai) and a Machine Learning Engineer at Microsoft where he serves on the CSE AI Ethics Review Board. His research focuses on applied technical and policy methods to address ethical, safety and inclusivity concerns in using AI in different domains. He has built the largest community driven, public consultation group on AI Ethics in the world that has made significant contributions to the Montreal Declaration for Responsible AI, the G7 AI Summit, AHRC and WEF Responsible Innovation framework and the European Commission Trustworthy AI Guidelines. His work on public competence building in AI Ethics has been recognized by governments from North America, Europe, Asia and Oceania. More information on his work can be found at https://atg-abhishek.github.io

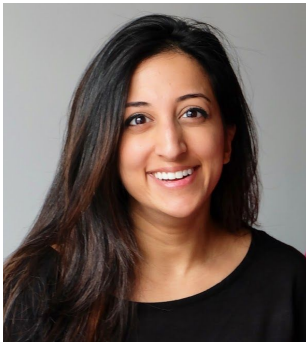

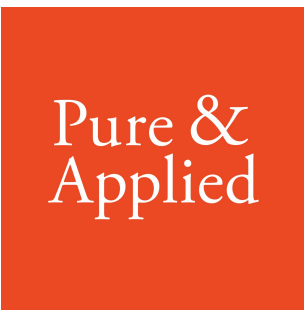

**Nikitasha (Niki) Kapoor** is the Principal Researcher at Pure & Applied Group, a research and learning design firm supporting the development of community-led programs and social innovations (www.pureandapplied.group). Whether conducting research to create solutions for complex organizational shifts or developing country-wide Indigenous-led social innovation programs across Canada, Niki brings both an analytic and creative lens to her work. This approach has enabled her to develop compelling and effective solutions, aligning people on common objectives, shared values, practices and goals. Motivated to create a future where diversity, equity and inclusion are embedded and operationalized in community development, Niki embraces ambitious mandates with confidence by fostering cross-sectoral partnerships and collaborating. Niki is multilingual and speaks English, French, Hindi, Punjabi and Urdu.